\documentclass[twocolumn,english,aps,prl,showpacs,superscriptaddress]{revtex4}

\usepackage[T1]{fontenc}
\usepackage[latin1]{inputenc}
\usepackage{graphicx}
\usepackage{hyperref}
\usepackage{babel}
\usepackage{amsfonts}

\begin{document}

\title{Bad-metal behavior reveals Mott quantum criticality in doped Hubbard models}
\author{J. Vu\v{c}i\v{c}evi\'{c}}

\affiliation{Scientific Computing Laboratory, Institute of Physics Belgrade, University
of Belgrade, Pregrevica 118, 11080 Belgrade, Serbia.}

\author{D. Tanaskovi\'{c}}

\affiliation{Scientific Computing Laboratory, Institute of Physics Belgrade, University
of Belgrade, Pregrevica 118, 11080 Belgrade, Serbia.} 

\author{M. J. Rozenberg}

\affiliation{Laboratoire de Physique des Solides, CNRS-UMR8502, Universit\'{e} de Paris-Sud, Orsay 91405, France.}

\author{V. Dobrosavljevi\'{c}}

\affiliation{Department of Physics and National High Magnetic Field Laboratory,
Florida State University, Tallahassee, Florida 32306, USA.}

\begin{abstract}
Bad-Metal (BM) behavior featuring linear temperature dependence of the resistivity extending to well above the Mott-Ioffe-Regel (MIR) limit is often viewed as one of the key unresolved signatures of strong correlation. 
Here we associate the BM behavior with the Mott quantum criticality by examining a fully frustrated Hubbard model where all long-range magnetic orders are suppressed, and the Mott problem can be rigorously solved through Dynamical Mean-Field Theory. We show that for the doped Mott insulator regime, the coexistence dome and the associated first-order Mott metal-insulator transition are confined to extremely low temperatures, while clear signatures of Mott quantum criticality emerge across much of the phase diagram. Remarkable scaling behavior is identified for the entire family of resistivity curves, with a quantum critical region covering the entire BM regime, providing not only insight, but also quantitative understanding around the MIR limit, in agreement with the available experiments.
\end{abstract}

\pacs{71.27.+a,71.30.+h}

\maketitle

Metallic transport  inconsistent with Fermi liquid theory has been observed in many different systems; it is often linked to quantum criticality around some ordering phase transition \cite{Sachdev_book,Vojta2007}. Such behavior is notable near quantum critical points in good conductors, for example in heavy fermion compounds \cite{stewartrev1,Stewart2001}. In several other classes of materials, however, much more dramatic departures form conventional metallic behavior are clearly observed, where resistivity still rises linearly with temperature, but it reaches paradoxically large values, well past the MIR limit \cite{GunnarssonRMP2003,Hussey2004}. This "Bad-Metal" (BM) behavior \cite{Emery1995}, was first identified in the heyday of high-temperature superconductivity, in materials such as $\mathrm{La}_{2-x}\mathrm{Sr}_x \mathrm{CuO}_4$ \cite{Takagi1992}. While the specific copper-oxide family and related high-$T_c$ materials remain ill-understood and marred with controversy, it soon became clear that BM behavior is a much more general feature  \cite{Hussey2004} of materials close to the Mott metal-insulator transition (MIT) \cite{Dobrosavljevic_book2012}. Indeed, it has been clearly identified also in  various oxides \cite{QazilbashPRB2006,maeno1998prb}, organic Mott systems \cite{LimelettePRL2003,Kagawa2005,Merino2008}, as well as more recently discovered families of iron pnictides \cite{Qazilbash2009natphys}. Despite years of speculation and debate, so far its clear physical  interpretation has not been established. 

To gain reliable insight into the origin of BM behavior, it is useful to examine an exactly solvable model system, where one can suppress all possible effect associated with the approach to some broken symmetry phase, or those specific to low dimensions and a given lattice structure. This can be achieved by focusing on the "maximally frustrated Hubbard model", where an exact solution can be obtained by solving Dynamical Mean-Field Theory (DMFT) equations \cite{ Georges1996} in the paramagnetic phase. Although various aspects of the DMFT equation have been studied for more than twenty years, only very recent work  \cite{Terletska2011,Vucicevic2013} established how to identify the quantum critical (QC) behavior associated with the interaction-driven Mott transition at half-filling. 

Here we present a large-scale computational study across the entire phase diagram, showing that qualitatively different transport behavior is found in doped Mott insulators. It reveals a clear and quantitative connection between BM phenomenology and the signatures of Mott quantum criticality, including the characteristic "mirror symmetry"  \cite{Dobrosavljevic1997} of the relevant scaling function. We demonstrate that the associated QC region, featuring linear temperature dependence of resistivity around the MIR limit, corresponds to a fully incoherent transport regime. In contrast, the coherent Fermi liquid (FL) regime and even the "Resilient Quasiparticle" regime \cite{CamjayiPRB2006,Deng2012} do emerge at lower temperature, but here the resistivity remains well below the MIR limit. Our results provide strong evidence that Bad Metallic behavior represents a universal feature of high temperature transport close to the Mott transition, presenting intriguing parallels with recent ideas based on holographic duality \cite{Aristomenis2013,hartnoll2014preprint}.

\begin{figure}[t]
\includegraphics  [width=3.2in]{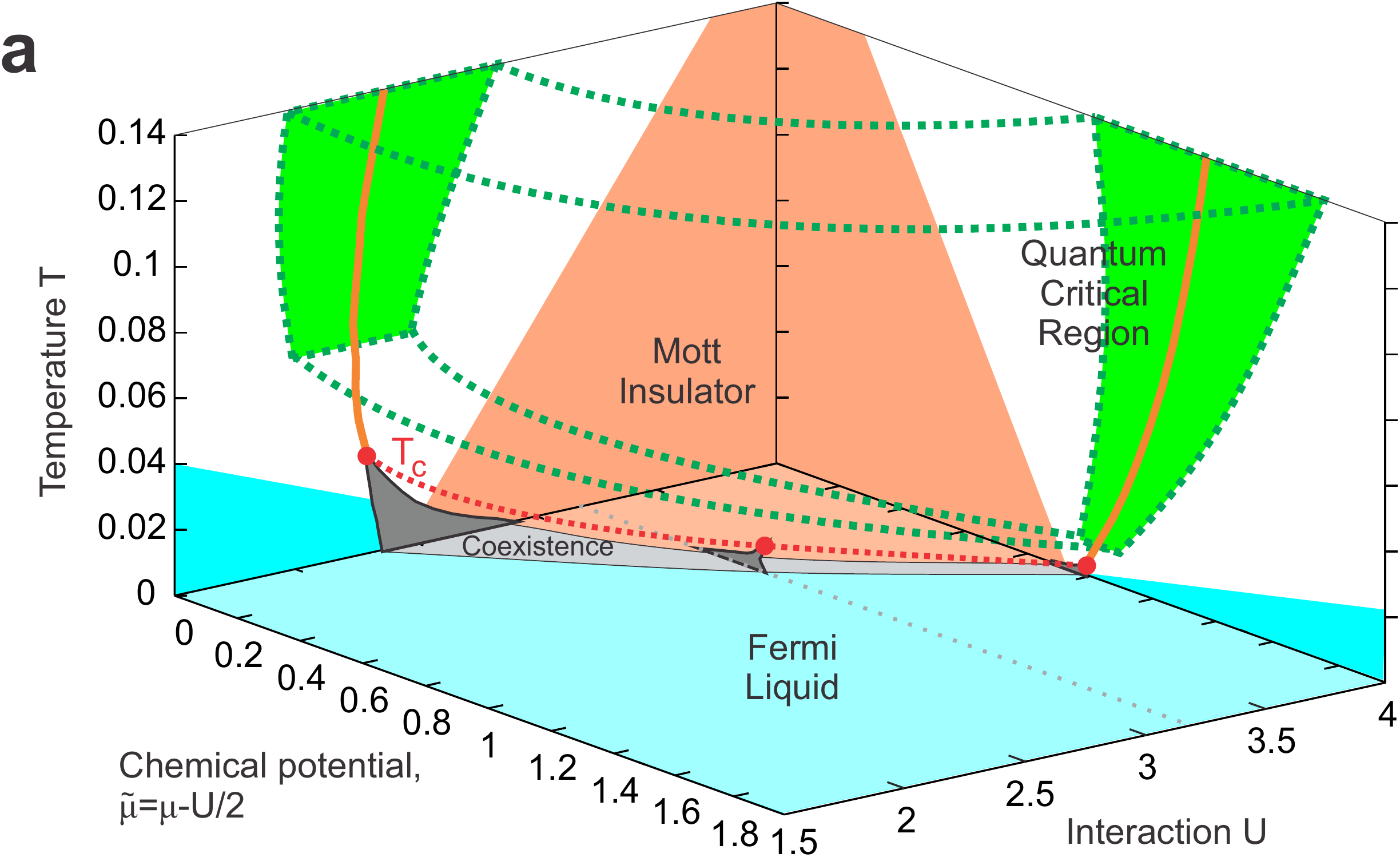}
\par
\vspace*{0.8cm}
\includegraphics  [width=3.2in]{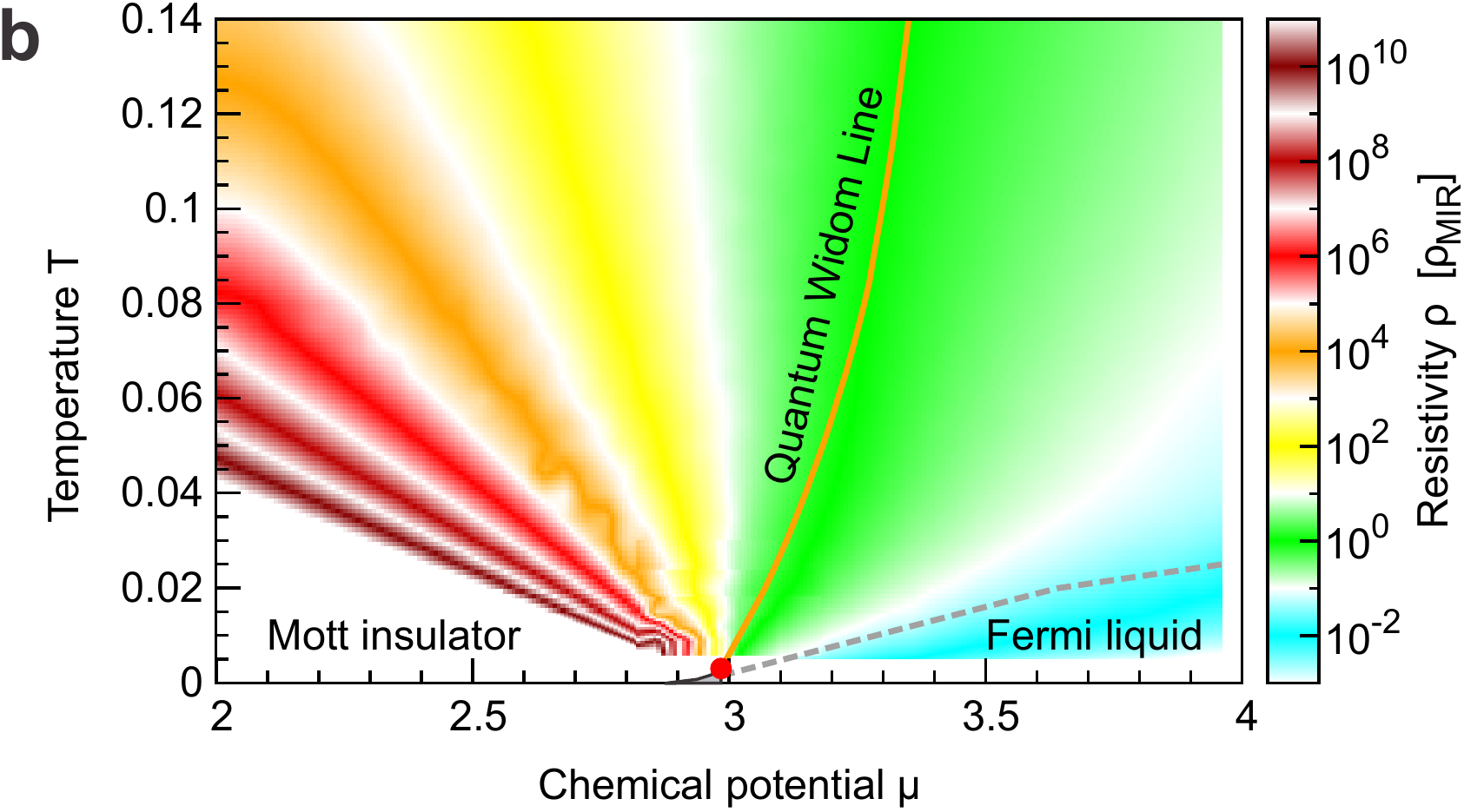}
\caption{
(a) Phase diagram of the maximally frustrated Hubbard model. 
The quantum critical scaling is observed in the green region which extends to lower temperatures as $T_c$ (red dots) is reduced. (b) Color plot of the resistivity in the $\mu -T$ plane for $U=4$. The quantum Widom line (see text) passes through the crossover region where the resistivity is around the MIR limit. The coexistence region (gray) is barely visible on the scale of this plot. }
\label{fig_phase_diagram_3D}
\vspace*{-.3cm}
\end{figure}

{\it Phase diagram.}
We consider a single-band Hubbard model defined by the Hamiltonian
$$
H =  - t \sum_{\langle i,j\rangle,\sigma} \left( c^\dagger_{i\sigma} c_{j\sigma} + \mathrm{h.c.}\right) 
           + U\sum_i n_{i\uparrow}n_{i\downarrow}
           -\mu\sum_{i,\sigma} c^\dagger_{i\sigma} c_{i\sigma},
$$
where $t$ stands for the nearest-neighbor hopping amplitude, $U$ is the on-site interaction, and $\mu$ denotes the chemical potential. The creation and annihilation operators for spin orientation $\sigma$ are denoted by $c^\dagger_{\sigma}$ and $c_{\sigma}$, and $n_{i\sigma}=c^\dagger_{i\sigma} c_{i\sigma}$. 
We solve the DMFT equations using the continuous time quantum Monte Carlo (CTQMC) algorithm for the impurity solver \cite{Werner2006,Haule2007,GullRMP2011}. We focus on the paramagnetic solution which is a physically justified assumption for frustrated lattices. We use the semi-elliptic bare density of states and set the half-bandwidth $D=1$ as the unit of energy. This corresponds to the infinitely-dimensional Bethe lattice, as well as the fully connected lattice with random hopping amplitudes \cite{Georges1996}.

At half-filling, strong enough on-site interaction $U$ opens a spectral gap at the Fermi level and produces the Mott insulating state \cite{Georges1996}. The Mott insulator can also be destroyed by adding electrons to the system, i.e raising the chemical potential $\mu$. When $\mu$ reaches the upper Hubbard band, the system is once again conducting \cite{CamjayiPRB2006}. 
In both cases, at low-temperature the transition is of the first order, and features a pronounced jump in the value of resistivity and other quantities \cite{Rozenberg2002}. Around the first order transition line, a small coexistence region is present, where both metallic and insulating phases are locally stable. Our calculations show (see Supplementary Sections I and II) that the critical  end-point temperature $T_c (U)$ for the doping-driven transition rapidly drops with increasing interaction, and at $U=4$ it already is less than $10\%$ of that at half-filling. This is illustrated in Fig.~\ref{fig_phase_diagram_3D}a. 
At the critical end-point (red dots) the two solutions merge, and above it no true distinction between the phases exists; only a rapid crossover is observed upon variation of $U$ or $\mu$. Previous work \cite{Terletska2011,Vucicevic2013} examined the vicinity of the interaction-driven MIT at half-filling; here we analyze the broad finite temperature crossover region between the half-filled Mott insulator and the doped Fermi liquid state \cite{Rozenberg2002,Zitko2013,Werner2007,AmaricciPRL2007}. This "Bad Metal" regime, displaying very different transport behavior than that found at half-filling, is the main focus of this work.

In Fig.~\ref{fig_phase_diagram_3D}b, we color-code the resistivity in the $(\mu,T)$ plane, calculated for  $U=4$. The resistivity is given in units of the Mott-Ioffe-Regel limit $\rho_{_{\mathrm{MIR}}}$ which is defined as the highest possible resistivity in a Boltzmann semi-classical metal, corresponding to the scattering length of one lattice spacing. Numerical value for $\rho_{_{\mathrm{MIR}}}$ is taken consistently with Ref.~\cite{Deng2012}. At $\mu=U/2$ the system is half-filled. At approximately $\mu=U-D=3$, the Fermi level enters the upper Hubbard band, and a first-order doping-driven MIT is observed at temperatures below $T_c=0.003D$. 
While the chemical potential is within the gap, 
a clear activation behavior, $\rho\sim e^{E_g/T}$, is found at low temperatures.
On the metallic side of the MIT, due to the strong electron-electron scattering, the resistivity grows rapidly with temperature, and typical Fermi-liquid behavior is observed only below rather low coherence temperature $T_{\mathrm{FL}}$ (denoted with the gray dashed line). 

\vspace*{.2cm}

{\it Quantum critical scaling.} 
In the standard scenario for quantum criticality \cite{Sachdev_book,Dobrosavljevic_book2012}, the system undergoes a zero-temperature phase transition at a critical value of some control parameter $g=g_c$, and within a ``V-shaped" finite temperature region, physical quantities display scaling behavior of the form $A(g,T) = A_c(T)\; F(T/(g-g_c)^{z\nu})$. 
Mott MIT is a first order phase transition \cite{Nozieres1998}, but the corresponding coexistence region is confined to extremely low temperatures, and at temperatures  sufficiently above the critical end-point $T_c$, the quantum effects are expected to set in \cite{Sachdev_book}, and restore the QC behavior. 

To test the QC scaling hypothesis in the case of a Mott transition, one must first identify the appropriate $g_c (T)$ "instability trajectory" \cite{Terletska2011,Vucicevic2013} which enters the argument of the scaling function (for illustration see Supplementary Fig.~2). 
It marks,  on the phase diagram, a trajectory where the system is least stable (i.e. is found in "equal proximity" to both the metal and the insulator), and is therefore most prone to fluctuations. The relevant thermodynamic stability is most easily determined from the curvature $\lambda$ of the free energy functional ${\cal{F}}[G(i\omega_n)]$ near its global minimum; this can be numerically determined by monitoring the convergence rate in the DMFT self-consistency loop \cite{Terletska2011}. Having in mind the analogy of this definition with the standard Widom crossover line for classical liquid-gas transitions \cite{Simeoni2010}, we refer to the instability line as the "quantum Widom line" (QWL) \cite{Vucicevic2013}.

\begin{figure}[t!]
\includegraphics  [width=3.4in]{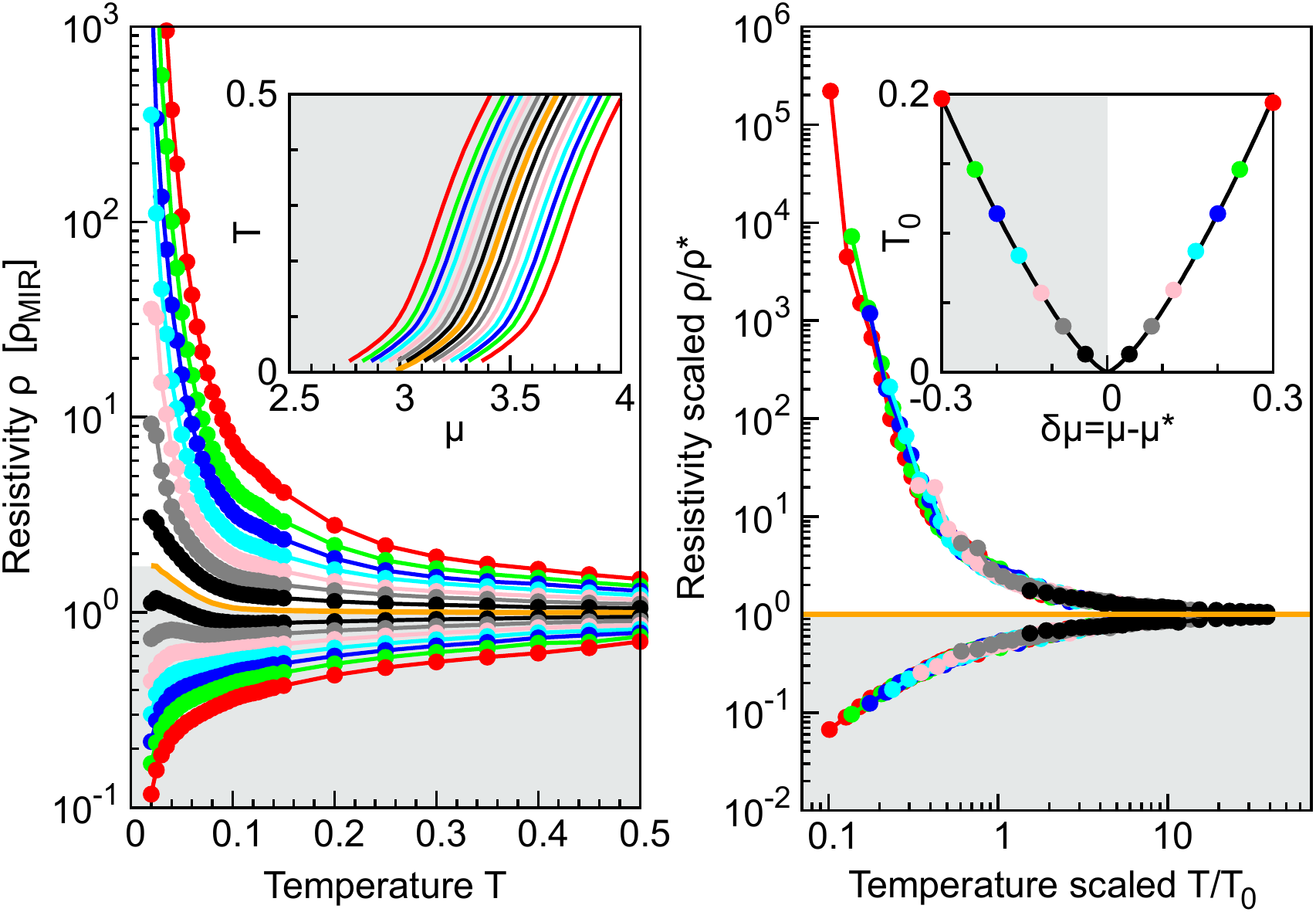}
\caption{
(a) Family of resistivity curves calculated along lines parallel to the QWL (orange). 
(b) Upon rescaling the temperature with adequately chosen parameter $T_0$, the resistivity curves collapse and reveal mirror symmetry of metallic-like and insulating-like behavior around the QWL.
$T_0$ depends on the distance from the QWL as $T_o (d\mu)\sim d\mu^{z\nu}$, with $z\nu \approx 1.35$.
}
\label{fig_mu_scaling}
\end{figure}

We carried out a careful {\it $\lambda$-analysis} for the doped Mott insulator (see Supplementary Section III), and we display the resulting QWL trajectory $\mu^*(T)$ as an orange line in all plots (throughout this Letter, an asterisk in the superscript indicates physical quantities evaluated along the QWL; e.g. $\rho^*(T)$ is resistivity calculated at temperature $T$ at $\mu = \mu^* (T)$). The QC region (green) spreads above the critical end-point (red points and dotted line) and quickly extends to much lower temperatures as $T_c$ is reduced (Fig.~\ref{fig_phase_diagram_3D}a). 
The QWL, separating the metallic-like and the insulating-like behavior, marks the center of the corresponding QC region, where the resistivity curves are expected to display the scaling behavior of the form
\begin{equation}\label{mu_scaling}
\rho(\mu,T)= \rho^*(T) F(T/T_0(d\mu )).
\end{equation} 
Here the parameter $T_0$ should assume power law dependence on the deviation from the QWL: $T_o (d\mu)\sim d\mu^{z\nu}$, with $d\mu = \mu-\mu^*(T)$. 

To check validity of the scaling hypothesis Eq.~(\ref{mu_scaling}), we calculate the resistivity along the lines parallel to the QWL,
as shown in Fig.~\ref{fig_mu_scaling}a. We find  that,  for the doped Mott insulator,  the resistivity shows very weak temperature dependence along the QWL. In particular, above $T=0.08$ it follows the line of constant resistivity {\em which coincides} with the MIR limit, $\rho^*(T > 0.08)=\rho_{_{\mathrm{MIR}}}$
(in contrast to the behavior previously established at half-filling \cite{Terletska2011,Vucicevic2013} where $\rho \gg \rho_{_{\mathrm{MIR}}}$ along the QWL).
In fact, all curves converge {\em precisely} to the MIR limit at high temperatures, suggesting its fundamental role in characterizing the metal-insulator crossover for doped Mott insulators. The curves also display the characteristic "bifurcation" upon reducing temperature, and a clear change in trend upon crossing the QWL.  
The scaling analysis confirms that all the curves indeed display fundamentally the same functional dependence on temperature, and that they all can be collapsed onto two distinct branches of the corresponding scaling function (Fig.~\ref{fig_mu_scaling}b). The scaling exponent has been estimated to be $z\nu\approx1.35\pm0.1$ for both branches of the scaling function, which display mirror-symmetry \cite{Dobrosavljevic1997,Terletska2011} over almost two decades in $T/T_0$, and the scaling covers more than three orders of magnitude in resistivity.

\vspace*{.2cm}

{\it Bad-Metal behavior.} 
We demonstrated the emergence of clearly defined quantum critical behavior thorough an analysis of the $(\mu,T)$ phase diagram, with $d \mu = \mu - \mu^*$ as the scaling parameter. From the experimental point of view it is, however, crucial to identify the corresponding QC region in the $(\delta,T)$ plane and understand its implications for the form of the resistivity curves for fixed level of doping $\rho(T)|_\delta$. By performing a careful calculation of the $\delta (\mu , T)$ dependence (see Supplementary Fig.~4), it is straightforward to re-plot our phase diagram and resistivity curves in the $(\delta , T)$ plane. Remarkably, we find that the quantum critical scaling region covers a broad range of temperatures and dopings, and almost perfectly matches the region of the well-known bad metal transport \cite{Pruschke1993,Deng2012}, characterized by the absence of long-lived quasiparticles and linear $\rho(T)|_\delta$ curves. We first analyze the $(\delta,T)$ phase diagram in detail, and then establish a connection between the slope of $\rho(T)|_\delta$ curves in the bad metal regime and the QC scaling exponent $\nu z$.

In Fig.~\ref{fig_delta_scaling}a we show the phase diagram of the doped Mott insulator. At $T=0$, the Mott insulator phase is found exclusively at zero doping. 
At low enough temperature and finite doping, characteristic Fermi liquid behavior is always observed. Here, the resistivity is quadratic in temperature, while a clear Drude peak is observed at low frequencies in optical conductivity and density of states  (see Supplementary Fig.~5). The coherence temperature $T_{\mathrm{FL}}$ is found to be proportional to the amount of doping $\delta$, however with a small pre-factor of about $0.1$, in agreement with Refs.~\cite{CamjayiPRB2006,Deng2012}. In a certain temperature range above $T_{\mathrm{FL}}$, a Drude peak is still present as well as the quasiparticle resonance in the single-particle density of states, but the resistivity no longer follows the FL $T^2$ dependence. This corresponds to the ``Resilient Quasiparticle'' (RQP) transport regime, which was carefully examined in Ref.~\cite{Deng2012}. At even higher temperatures, the temperature-dependent resistivity at fixed doping $\rho(T)|_\delta$ enters a prolonged linear regime (see Fig.~\ref{fig_delta_scaling}b) {\cite{JarrellPRB1994}, which is accompanied by the eventual disappearance of the Drude peak around the MIR limit. This behavior is usually referred to as the Bad Metal regime \cite{Deng2012}.  
The resistivity is comparable to the MIR limit throughout the BM region, and the QWL (as determined from our thermodynamic analysis) passes through its middle. 

\begin{figure}[t!]
\includegraphics  [width=2.5in, page=1, trim=0.0cm 0.0cm 0.0cm 0.0cm]{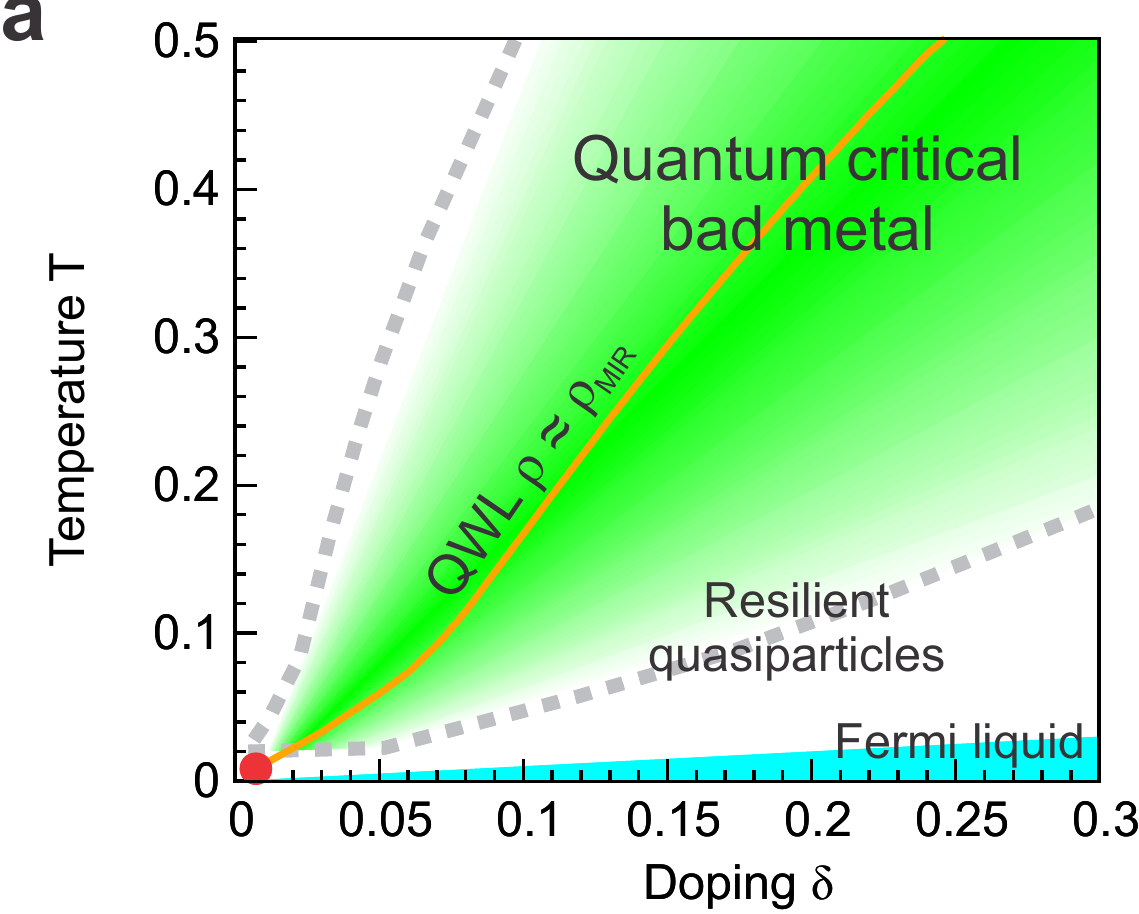}
\par 
\vspace{0.5cm}
\includegraphics  [width=2.5in, page=1]{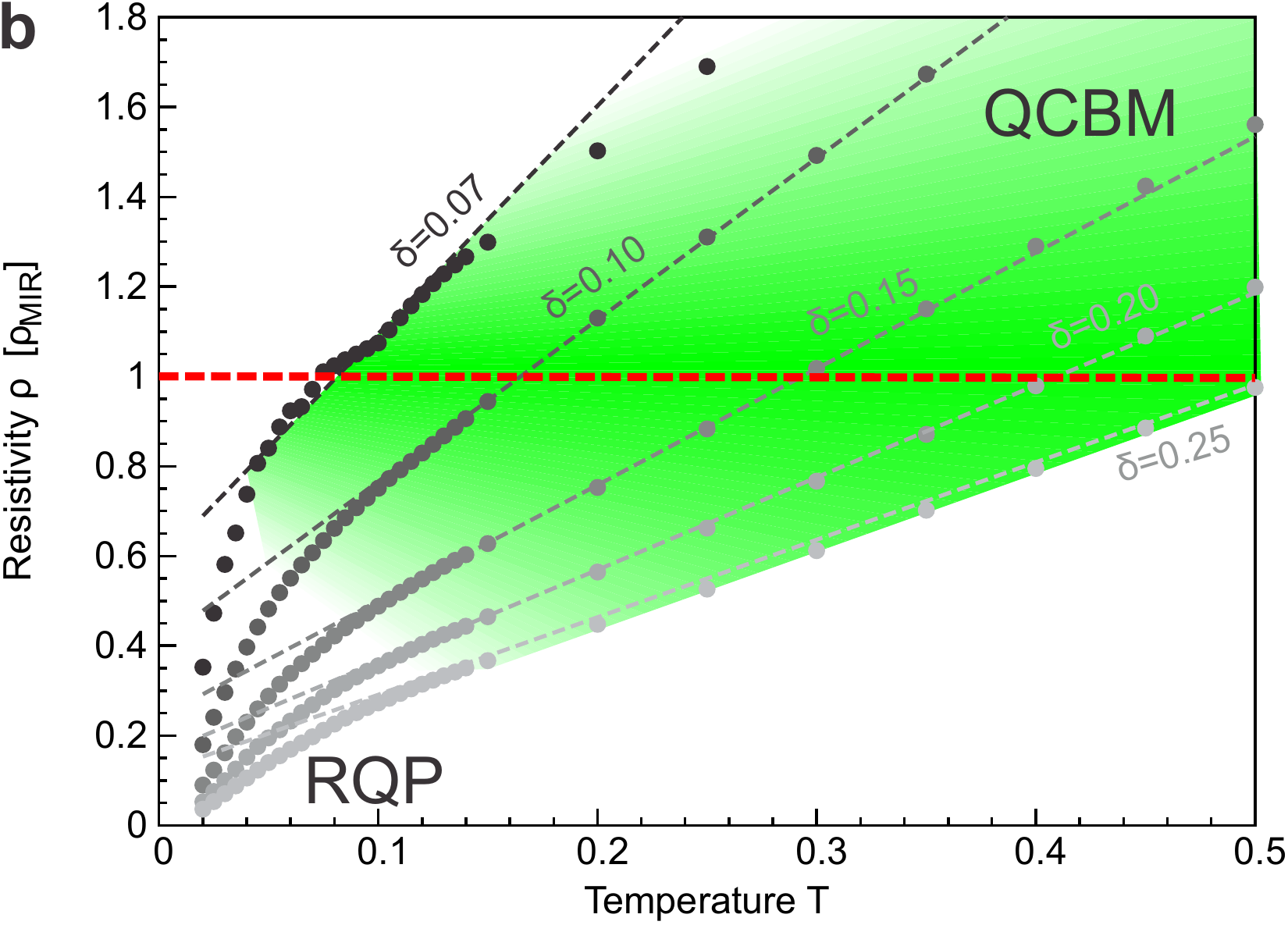}
\caption{ (a) DMFT phase diagram of the doped Mott insulator on a frustrated lattice. The bad metal (green) region matches perfectly the region of quantum critical scaling. (b) The bad metal regime features linear temperature dependence of resistivity with the slope roughly proportional to an inverse power law of doping which we find to be a consequence of underlying quantum criticality. 
}
\label{fig_delta_scaling}
\end{figure}

The region of linear $\rho(T)|_\delta$ dependence is found to be completely encompassed by the QC scaling region between the dashed lines on Fig.~\ref{fig_delta_scaling}a (see Supplementary Section VI).
We therefore expect that the emergence of the linear-T dependence of the resistivity, as well as the doping dependence of its slope, should be directly related to the precise form of the corresponding scaling function. Indeed, at high temperature and close to the QWL, the argument of the scaling function $x=d \mu/T^{1/z\nu}$ is always small, and the scaling function can be linearized, viz. $\tilde{F}(x)\approx 1+Ax+\cdots $. We find that the coefficient $A$ has the numerical value $A\approx - 0.74$. The functional form for $\rho(T)|_\delta$ close to the QWL is then directly determined by the behavior of the scaling parameter $x(T)|_\delta$. We find  that $x(T)|_\delta$ is a linear function in a wide range of temperature around $T^*(\delta)$. Then, close to the QWL, the resistivity is well approximated by a linear function of the form
\begin{eqnarray}\label{dopingdependence}
\rho(T)|_\delta &\approx& \rho^*(\delta) \left(1+A\left.\frac{\partial x}{\partial T}\right|_{\delta,T=T^*(\delta)}(T-T^*(\delta))\right) .
\end{eqnarray}
Furthermore, the slope of the scaling argument at the QWL can be expressed as
$ \left.\frac{\partial x}{\partial T}\right|_{\delta,T=T^*(\delta)} = \left(\chi^*(\delta) \frac{dT^*}{d\delta} (T^*(\delta))^{1/z\nu}\right)^{-1} $, where 
$\chi^*(\delta)=\frac{\partial \delta}{\partial \mu}|_{T=T^*(\delta)}$.
Here we observe that the charge compressibility is nearly constant along the QWL, $\chi^*(\delta) \approx \chi^*=0.33$, see Supplementary Fig.~6, which may be interpreted as another manifestation of the quantum critical behavior we identified. $T^*(\delta)$ is approximately a linear function $T^*(\delta) \approx K_0 + K \delta$, where $K \approx 2$ and $K_0$ is small. In Fig.~\ref{fig_delta_scaling}b we compare the approximation stated in Eq.~(\ref{dopingdependence}) with the DMFT result and find excellent agreement. 

\begin{figure}[t]
\includegraphics  [width=4.5in, trim=4.0cm 6.0cm 1.0cm 9.0cm]{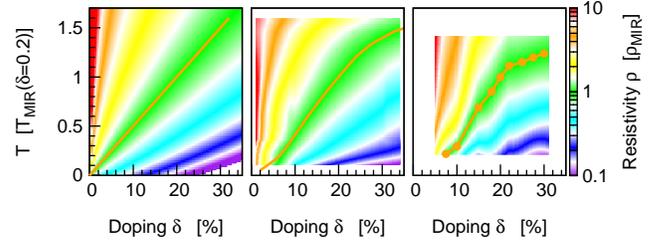} 
\caption{ Resistivity given by (a) the semi-analytical formula obtained from the scaling hypothesis, (b) DMFT result, and (c) the experimental result on cuprate $\mathrm{La}_{2-x}\mathrm{Sr}_x \mathrm{CuO}_4$ samples from Ref.~\onlinecite{Takagi1992SN}.}
\label{fig_takagi}
\end{figure}

Finally, noting that for $\delta>5\%$, $\rho^*(\delta)=\rho_{_{\mathrm{MIR}}}$, we arrive at the central result of this Letter:
\begin{equation}\label{dopingdependence2}
\rho_{_{\mathrm{QCBM}}}(T)|_\delta \approx \rho_{_{\mathrm{MIR}}} \left(1+C \, \delta^{-1/z\nu}(T-K\delta)\right).
\end{equation}
In the quantum critical bad metal regime, the resistivity has a linear temperature dependence with the slope decreasing as a power $-1/z\nu$ of doping. This demonstrates a direct connection of the universal high temperature behavior in the Bad Metal regime with the (zero-temperature) quantum phase transition. The MIR limit of the resistivity is reached at temperature roughly proportional to the amount of doping,  $T^*(\delta) \propto \delta$, since the doping level sets the main energy scale in the problem. The result of this simplified scaling formula is color-plotted in Fig.~\ref{fig_takagi}a (with $C=0.69$, $K=1.97$ and $z\nu=1.35$) and shown to capture the features of the full DMFT solution at high temperatures.

\vspace*{.2cm}

{\it Discussion.} Sufficiently systematic experimental studies of doped Mott insulators, covering an appreciable range of doping and temperature, remain relatively scarce. Still,  approximately linear temperature dependence of the resistivity at high temperatures with the slope that decreases with doping has been observed, most notably in the seminal work of  Takagi {\em et al.}  \cite{Takagi1992} on $\mathrm{La}_{2-x}\mathrm{Sr}_x \mathrm{CuO}_4$. To make a qualitative comparison with our theory and to highlight a universal link of Bad Metal behavior and quantum criticality associated with the Mott metal-insulator transition, in Fig.~\ref{fig_takagi} we color code the reported experimental data; here the temperature is shown in the units of $T_{_{MIR}}$ at $20\%$ doping and the resistivity is given in units of $\rho_{_{\mathrm{MIR}}}$, which  in this material is estimated as 1.7 ${\mathrm m} \Omega {\mathrm cm}$. The experimental results presented in Fig.~\ref{fig_takagi}c cover the temperature range of $150-1000$ K at $5-30\%$ doping. Here one observes  a striking similarity between DMFT theory and the experiment, as already noted in early studies \cite{JarrellPRB1994,Pruschke1993,Note_exponent}.  We established this result by focusing on an exactly solvable model, where all ordering tendencies are suppressed, and single-site DMFT becomes exact. Real materials, of course, exist in finite (low) dimensions where systematic corrections to DMFT need to be included \cite{JarrellRMP2005,KotliarRMP2006,Sordi2013,GullPRL2013}. In many cases \cite{Georges2011,Tanaskovic2011,Kokalj2013}, these nonlocal corrections prove significant only at sufficiently low temperatures. Then our findings should be even quantitatively accurate in the high-temperature incoherent regime, as in the very recent experiments on organic materials \cite{Furukawa2015} for the case of half-filling.

We thank E. Abrahams, H. Alloul, S. Hartnoll, N. Hussey, K. Kanoda,
A. Schofield, Q. Si, J. Schmalian and M. Vojta
for useful discussions. J.V. and D.T. acknowledge support
from the Serbian Ministry of Education and Science under project No. ON171017. V.D. was supported by the NSF grants DMR-1005751 and DMR-1410132.
Numerical simulations were run on the AEGIS e-Infrastructure, supported in part by FP7 projects EGI-InSPIRE and PRACE-3IP. J. V., D. T. and M. J. R. acknowledge support from the
bilateral French-Serbian PHC Pavle Savic 2012-2013 grant.

\bibliographystyle{apsrev}
\bibliography{BMQC}


\pagebreak

\setcounter{figure}{0} 
\setcounter{equation}{0} 

{\bf SUPPLEMENTARY INFORMATION}

\section{ I. Numerical details: the DMFT loop and the impurity solver}

We have used the CTQMC impurity solver as implemented by K.~Haule in Ref.~\onlinecite{Haule2007SN}. We have used $4-6\times 10^9$ Monte Carlo steps.
When $T>0.14$, the high frequency tail was calculated from the atomic limit and Hubbard-I approximation was used otherwise. At high temperatures, 10-15 DMFT iterations were usually sufficient to reach the self-consistent solution with the accuracy $|G^i(i\omega_0)-G^{i-1}(i\omega_0)|\approx10^{-4}$, where $\omega_0=\frac{1}{2}\pi T$. In the coexistence region, we used a larger number of DMFT iterations (up to 30) to test the stability of the obtained solution.

\section{II. Determination of $T_c$}

\begin{figure}[t]
\includegraphics  [width=3.2in, page=1, trim=2.5cm 2.0cm 2.0cm 2.0cm]{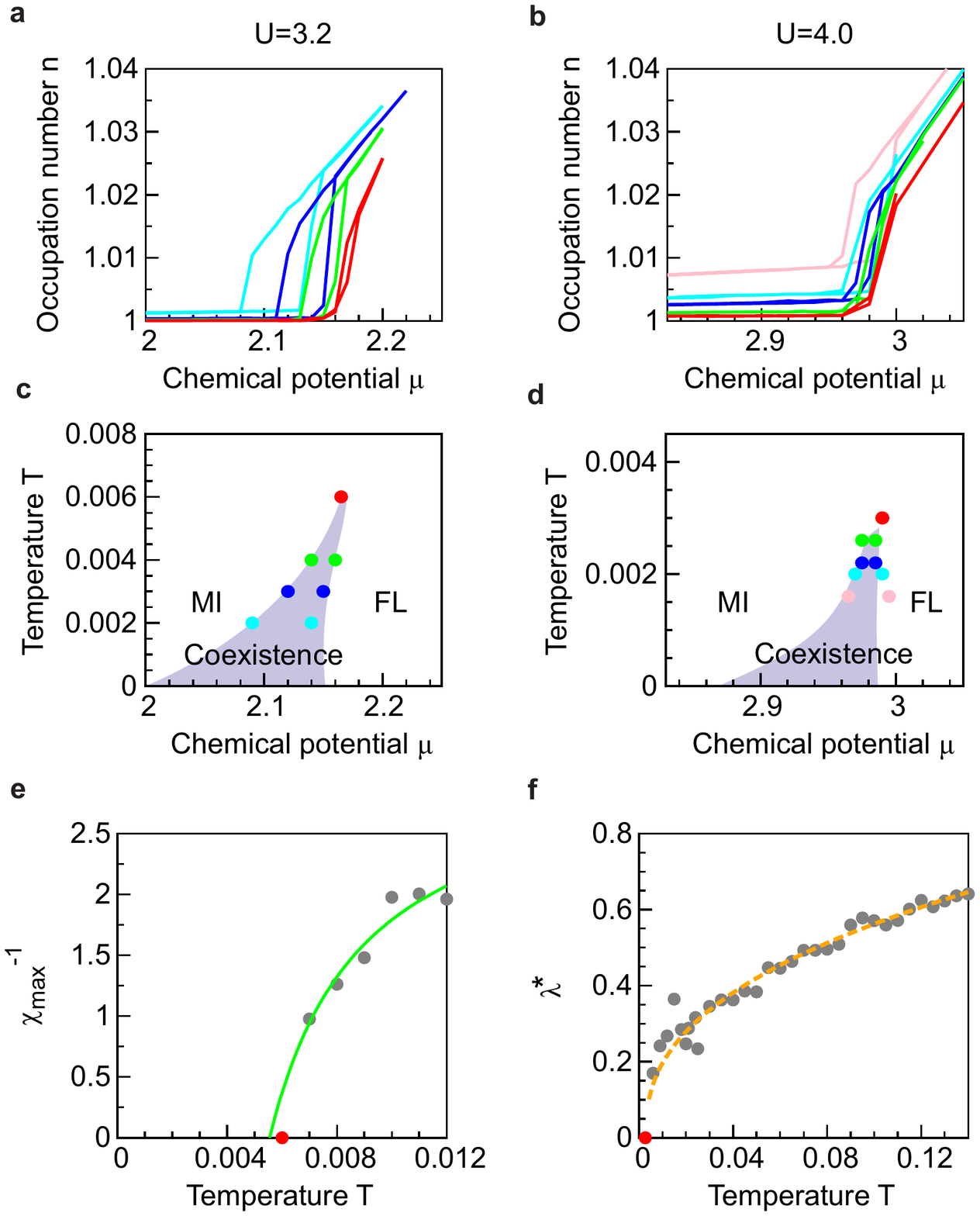}
\caption{ \textbf{ Coexistence region of the first-order doping-driven Mott metal-insulator transition can be determined in different ways.}  (a),(b),(c),(d) The position of spinodals can be determined from the jumps in the occupation number. In the coexistence region, two types of solution are possible, depending on the initial guess in the DMFT procedure. This is observed as a hysteresis loop in the occupation number and other quantities. (e),(f) Precisely at the critical point, physical quantities often have extremal values (zero or infinity). By extrapolating such quantities from higher temperatures, one can estimate the critical temperature. (e) The maxima of the inverse charge compressibility with respect to the chemical potential can be extrapolated to obtain a good estimate for $T_c$. (f) The values of $\lambda$ along the instability line $\mu^*(T)$ become scattered and overestimated close to the critical point, due to numerical error from the CTQMC. This makes it unpractical to to use extrapolation of $\lambda^*$ for estimation of $T_c$.}
\label{fig_coexistence}
\end{figure}

\subsection{$T_c$ from the position of the spinodals}
The first order phase transition is most easily observed by looking at the occupation number. At very low temperature, while the chemical potential is within the spectral gap, filling is roughly a constant, i.e. $n(\mu)\approx0.5$. When the chemical potential reaches the upper Hubbard band, a quasi particle peak forms abruptly at its lower edge causing an immediate transfer of spectral weight from the lower Hubbard band to the vicinity of the Fermi level \cite{Georges1996SN,Zitko2013SN}. This is observed as a jump in the occupancy from nearly half-filling to around 2-3\% doping. An insulating solution is not possible when $\mu$ is in the upper Hubbard band, hence its bottom edge determines the insulating (right) spinodal. However, a metallic solution is possible even when $\mu$ is in the gap. This type of state features an in-gap quasi-particle peak \cite{Fisher1995SN} and is observed in the coexistence region. The lowest value of the chemical potential at which the quasi particle peak can survive constitutes the metallic (left) spinodal. The disappearance of the QP peak at the metallic spinodal is also abrupt, and occurs at finite doping. Therefore, there is a range of doping that is not achievable locally at any value of the chemical potential, but only globally through phase separation. With increasing temperature, the forbidden doping range shrinks and disappear together with the hysteresis loop, precisely at $T_c$ \cite{Rozenberg2002SN,Werner2007SN}. Note also, that the range of forbidden doping vanishes at $T=0$ as well, where a metallic solution is possible even at infinitesimal doping \cite{Georges1996SN}, although in this case particle-hole symmetry is broken and $\mu\neq U/2$. In Supplementary Figure \ref{fig_coexistence}a,b we show the hysteresis curves of the occupancy for two values of interaction $U$. The position of spinodals and the width of the coexistence region are easily determined from the jumps in $n(\mu)$. We considered the lowest temperature at which no coexistence is observed to be the critical temperature. Note also that due to the numerical error of the CTQMC, some unphysical doping is observed in the insulating state at the lowest temperatures. We were not able to obtain physically meaningful results below $T\approx0.0015$ and this is the lowest temperature at which we have found the method to be reliable. The numerical error from the CTQMC becomes significant at low temperature and a precise assessment of $T_c$'s lower than $\approx0.002$ proves very difficult. 
The coexistence regions at the two values of $U$ are shown in Supplementary Figure \ref{fig_coexistence}c and d. The $T=0$ position of the left spinodal is taken from the ED calculation found in \cite{Fisher1995SN} and seems to fit well our finite temperature results. 

\subsection{$T_c$ from the charge compressibility}

The alternative way of determining $T_c$ is by looking at the uniform charge susceptibility $\chi=\frac{\partial n}{\partial \mu}$. Precisely at the critical point, $\chi$ is divergent and above $T_c$, there is a line of maxima in $\left.\chi(\mu)\right|_T$.  Furthermore, it can be shown \cite{Rozenberg1999SN} that close to the critical point $\chi^{-1}\sim \frac{T-T_c}{a+b(T-T_c)}$. This is useful as one can extrapolate the values of $\chi^{-1}_{\mathrm{max}}(T)$ to lower temperatures and see where it goes to zero. However, such method is of inferior accuracy compared to the direct observation of the coexistence, and we use it only for cross-checking of our results. In Fig.~\ref{fig_coexistence}e  we show such calculation in the case of $U=3.2$.

\subsection{$T_c$ from the $\lambda$ analysis}

In Supplementary Figure \ref{fig_coexistence}f we plot the values of $\lambda$ along the instability line (see the next Section). Close to the the critical point, it is very difficult to make a precise estimate of the DMFT convergence rate, as high convergence is not achievable at all. The low temperature values are therefore much more scattered and systematically overestimated. Although in principle one could estimate $T_c$ from higher temperatures by extrapolating $\lambda^*(T)\equiv\lambda(\mu^*(T),T)$, the numerical noise makes such a method very impractical. Further difficulty lies in the possibility of $\lambda^*(T)$ changing trend before going to zero, which introduces additional systematic error to the estimate of $T_c$.

\section{III. Quantum Widom Line and the $\lambda$ analysis}

In our previous work \cite{Vucicevic2013SN}, we have discussed a possible generalization of the Widom line (originally defined in the context of classical liquid-gas transitions \cite{Simeoni2010SN}), 
to strictly zero-temperature (quantum) phase transitions (see Supplementary Fig.~2). The most natural way of defining such a quantum Widom line is by looking at the free-energy landscape around the ground state of the system, as it is well defined in all physical models. Regardless of the specifics of the phase transition, precisely at the critical point, the free energy minimum is flat, i.e.~its curvature $\lambda$ is zero. At higher temperatures, this leads to a line of minima in $\lambda$ with respect to the parameter that is driving the transition (at half-filling we had $\left.\frac{\partial\lambda}{\partial U}\right|_T=0$). It is at those minima that the fluctuations are most pronounced - the system is ``equally close'' to the two competing phases and thus the least stable. Now we utilize this concept in the case of doping-driven Mott transition, and at each temperature search for the minimum value of $\lambda$ with respect to the chemical potential.

\begin{figure}[b!]
\includegraphics  [width=3.2in]{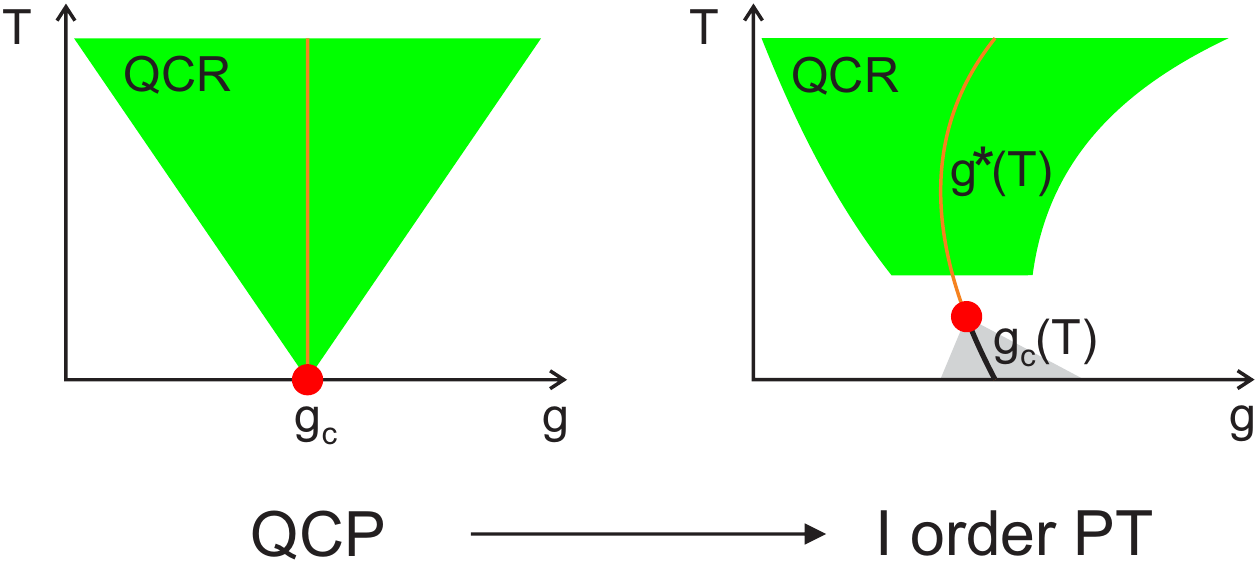}
\caption{\textbf{Standard QCP scenario is modified in the case of the Mott MIT.} At low temperature, the Mott MIT is of (weakly) first order  character, and features a coexistence region (gray) where both phases are locally stable. However, at temperatures sufficiently above the (very low) critical temperature $T_c$, the QC scaling holds (green). The critical transition-driving parameter $g_c$ is replaced by a more general, temperature dependent quantity. Below $T_c$ this is the line of first order transition $g_c(T)$ where the two states are equally favorable. Above $T_c$ it is the line of "maximal instability" of the ground state (see text), or the quantum Widom line $g^*(T)$. }
\label{fig_ilustracija}
\end{figure}

In practice, we calculate $\lambda$ by monitoring the convergence rate of the iterative DMFT procedure \cite{Terletska2011SN}. 
Given the model parameters, the free energy functional ${\cal F}_{U,T,\mu}[G(i\omega_n)]$ yields a smooth manifold in the Hilbert space of the Green's functions. Being Taylor expandable, the local environment of any free-energy minimum has to be parabolic. Thus, in the advanced stage of the DMFT procedure, i.e. close to the self-consistent solution, a steady, exponential convergence should be observed. The
curvature $\lambda$ is then directly related to the exponent of 
the functional dependence of the difference between the consecutive solutions versus the iteration index. When determining the convergence rate, however, it is not always sufficient to look at the Green's function in only the lowest Matsubara frequency, and one must use the generalized Raileigh-Ritz (RR) formula \cite{Case2007SN}
\begin{equation} \label{RRformula}
\lambda_i = 1 - \sum_n \frac{|G_n^{i+1}-G_n^{i}||G_n^{i}-G_n^{i-1}|}{|G_n^{i+1}-G_n^{i}|^2}, 
\end{equation}
where $i$ stands for the iteration index, and ideally, $\lambda = \lim_{i\rightarrow \infty} \lambda_i $.
However, the highest achievable level of convergence is determined by the amount of statistical noise in the CTQMC result, and when it is reached, $G(i\omega_n)$ just fluctuates around the self-consistent solution, and no further convergence is observed. Especially near the critical point, CTQMC error becomes substantial and a high convergence can not be reached at all. Here, typically only a few iterations are available for the estimation of $\lambda$ as most of the parabolic region is below the level of numerical noise, and one must look carefully for the range of iterations in which a steady exponential convergence is observed. 

The result presented with gray dots in Fig.~3a
is obtained by employing the RR formula from equation (\ref{RRformula}) at each iteration $i$,
and then taking the average over the set of 5 consecutive iterations that shows the least variance, i.e. the one corresponding to the period of the steadiest exponential convergence.

Away from half-filling, however, there are additional difficulties. Namely, $G(i\omega_n)$ is complex, which means that it has additional degrees of freedom as compared to its purely imaginary analogue at particle-hole symmetry. Thus, the fluctuations encountered in the convergence rate of $G(i\omega_n)$ are more severe, and the $\lambda$-analysis is harder to perform compared to the half-filled case.
This is why the data points presented with gray dots in Fig.~\ref{fig_instability_line} exhibit considerable scattering, but the overall trend is still rather obvious. In all of the calculations regarding the quantum critical (QC) scaling analysis, we use the smooth fit (orange dashed line) as the instability line and denote it with $\mu^*(T)$. Note that no other smoothing has been performed on the data, and all the minima are estimated automatically from the raw $\lambda$ results. Although there are considerable error bars on each $\mu^*(T)$ value, the high resolution in temperature increases the certainty of the result.

It is interesting that $\mu^*(T)$ is very close to the line of maxima of the second derivative of the occupation number versus the chemical potential, $\left.\frac{\partial^2 n}{\partial \mu^2}\right|_T=\mathrm{max}$. This is the place where $n(\mu)$ changes trend, and as expected, the instability line separates the metallic-like and insulating-like behavior on the phase diagram. 
Also note that $\mu^*(T)$ roughly follows an iso-resistive curve and so the resistivity does not change considerably along the instability line. At $T>0.08$ $\rho^*$ is found to be constant and equal to the Mott-Ioffe-Regel (MIR) limit. Above $T=0.14$, $\lambda$-analysis can not give reliable results as the minimum in $\lambda(\mu)|_T$ becomes very shallow, i.e. of depth comparable to the level of numerical noise. Throughout the paper we extrapolate the instability line to high temperatures $T>0.14$ by imposing the criterion $\rho^*=\rho_{\mathrm{MIR}}$. Also note that at very low temperature, the actual form of $\rho(\mu^*(T),T)$ depends strongly on the precise values of $\mu^*(T)$ because, in this region, the resistivity changes rapidly with the chemical potential.

\begin{figure}[t]
\includegraphics  [width=3.2in, trim=5.5cm 4.5cm 4.5cm 4.5cm]{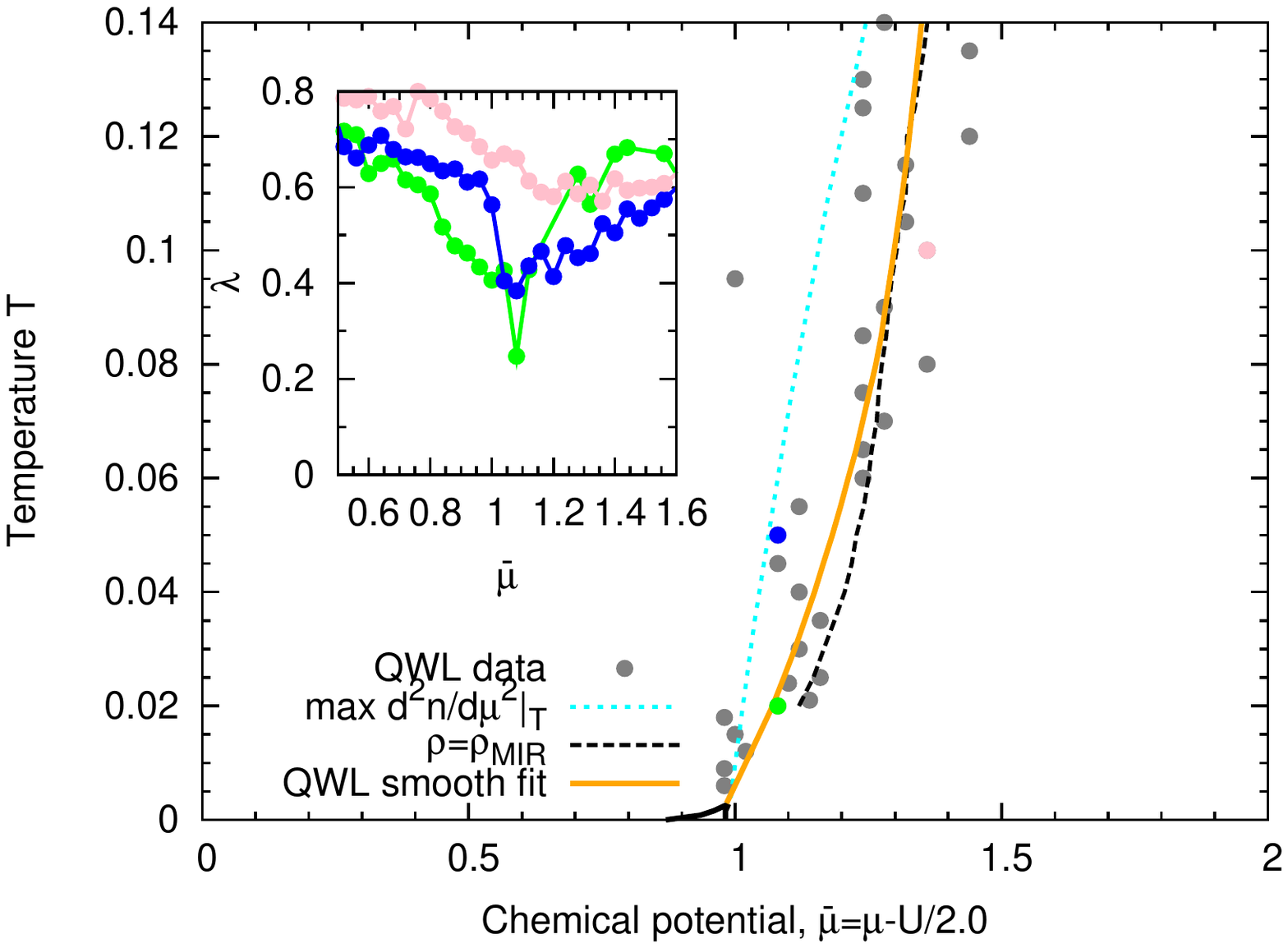}
\includegraphics  [width=3.2in, trim=5.5cm 4.5cm 4.5cm 4.5cm]{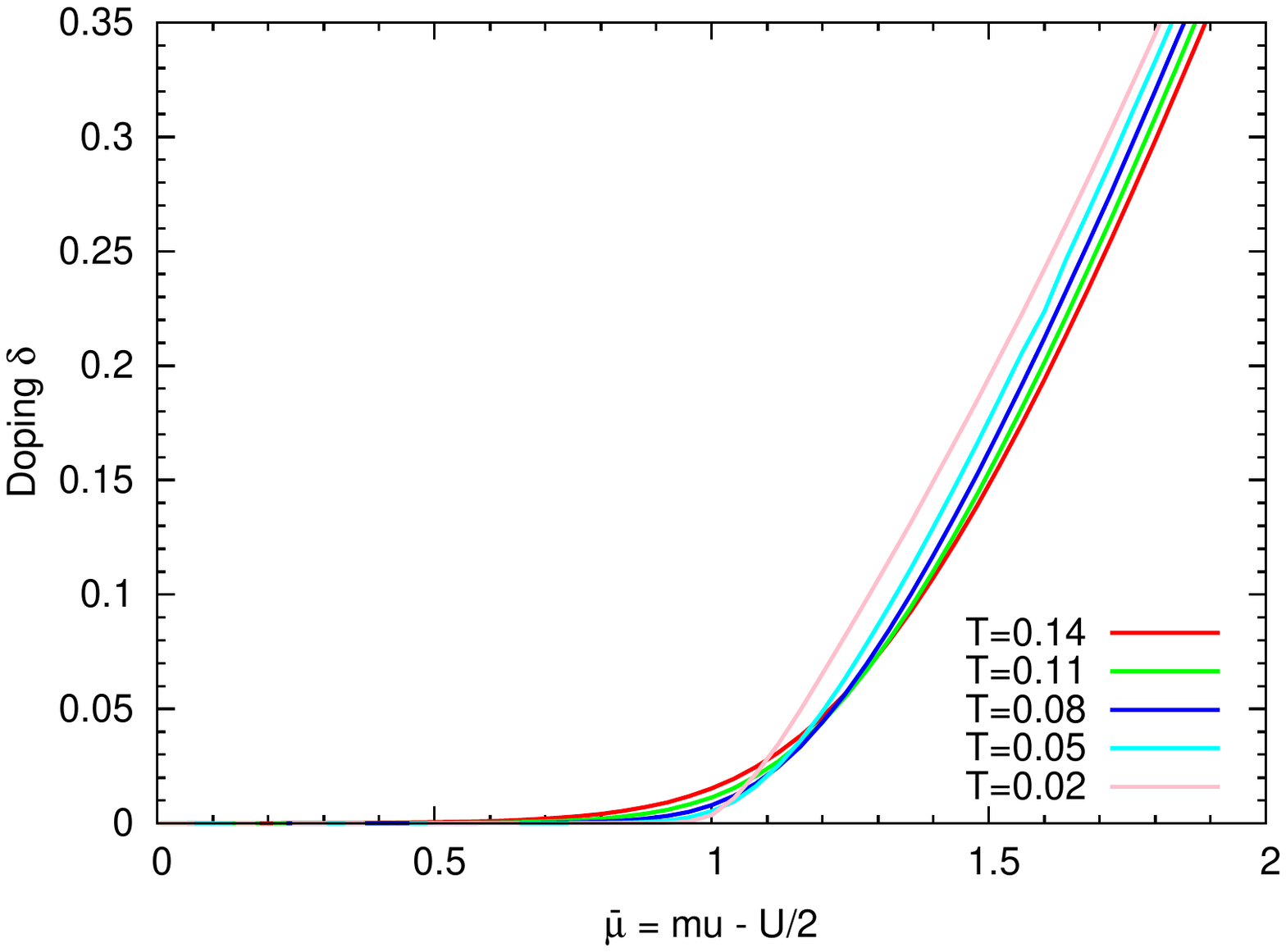}
\caption{\textbf{The instability line $\mu^*(T)$ (orange dashed) corresponds to the minima in $\lambda(\mu)|_T$, which is related to the convergence rate of the DMFT loop.} (a) The precision of $\lambda$ results is limited by the statistical noise in CTQMC. However, the minima in $\lambda(\mu)|_T$ are still clearly present, and $\mu^*(T)$ can be determined with satisfactory accuracy. At high temperature, QWL is found to coincide with the iso-resistive curve of the MIR limit (black dashed), which is then used to extrapolate the QWL to temperatures above $T=0.14$, where $\lambda$-analysis is no longer reliable. (b) The QWL is also very close to the point where occupancy $n(\mu)|_T$ changes trend, i.e. has a maximum of the second derivative. The line of maxima in $d^2n/d\mu^2|_T$ can also be considered a crossover line between metallic and insulating behavior (light blue dotted line on panel (a)). }
\label{fig_instability_line}
\end{figure}

\section{IV. Analytical continuation and calculation of resistivity}

\begin{figure}[b!]
\includegraphics  [width=3.2in]{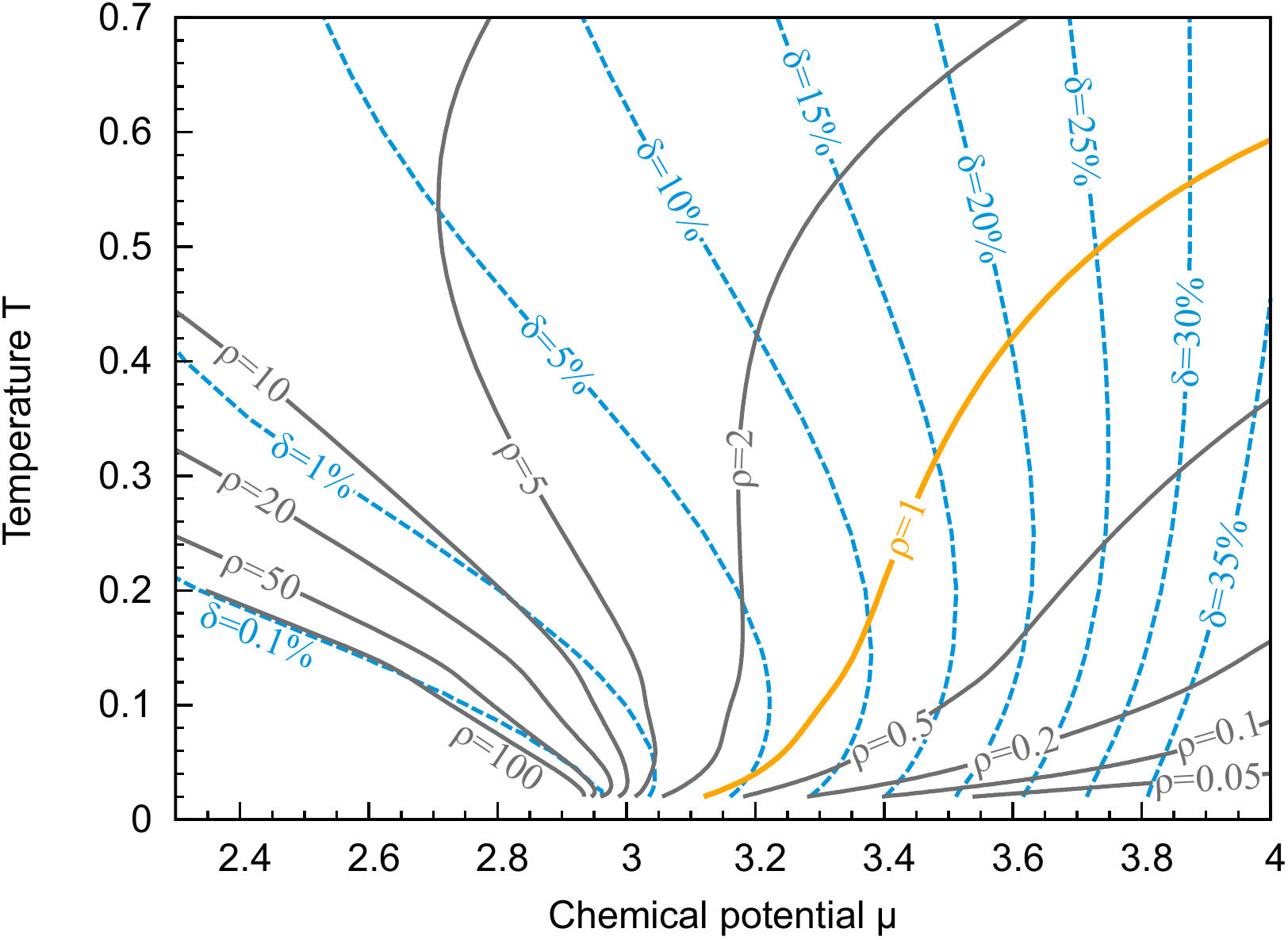}
\caption{\textbf{Lines of constant doping intersect with the QWL (orange), along which the resistivity is equal to the MIR limit  $\rho_{_{MIR}}$.}}
\label{fig_lines_and_doping}
\end{figure}

The straightforward application of the maximum entropy method (MEM) \cite{Jarrell1996SN,Sandvik1998SN} for analytical continuation of the Green's function can in some cases lead to unphysical results. In the metallic phase, this method tends to overestimate the height of the quasi-particle (QP) peak in the single-particle energy spectrum given by $-\frac{1}{\pi}\mathrm{Im} G(\omega+i0^+)$. Sometimes in those cases, the imaginary part of the self-energy falsely goes to zero at several frequencies (usually two or four), yielding an unphysical vanishing DC resistivity. Given the analytically continued Green's function on the real axis, the self-energy is obtained from the DMFT self-consistency condition
\begin{equation}
\Sigma(\omega)=\omega+\mu-G^{-1}(\omega)-t^2G(\omega),
\end{equation}
and the imaginary part of the above equation reads
\begin{equation}
\mathrm{Im}\Sigma(\omega) = \mathrm{Im}G(\omega)(|G(\omega)|^{-2}-t^2).
\end{equation}
It is immediately obvious that $|G(\omega)|=1/t$ yields $\mathrm{Im}\Sigma(\omega) = 0$, at any frequency. When there is an unphysical excess of QP weight, precisely this is seen, usually at the edges of the QP peak. This makes the conductivity integral divergent and the DC resistivity exactly zero. 

\begin{figure}[t!]
\includegraphics  [width=3.2in]{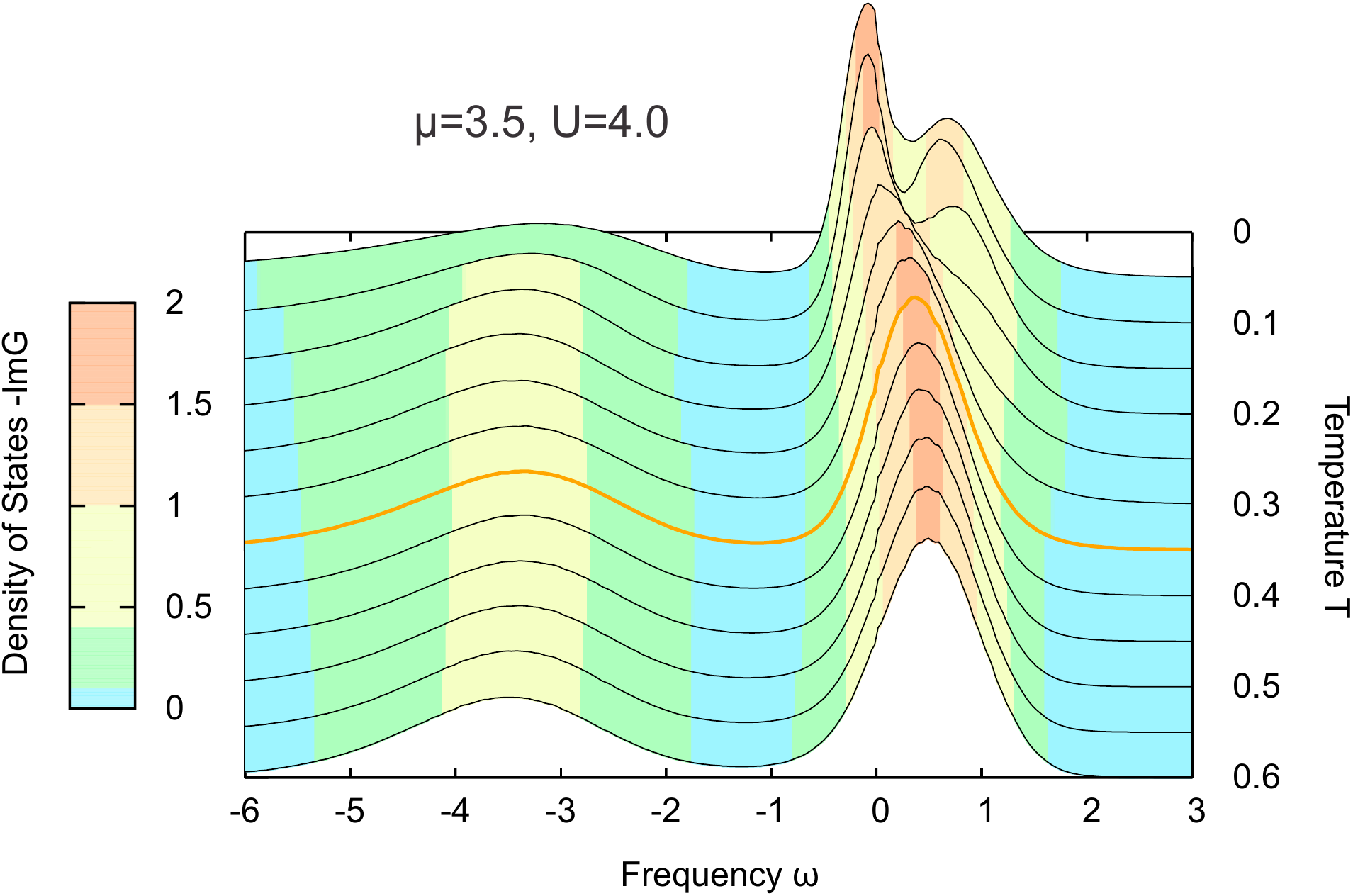}
\caption{\textbf{Evolution of the density of states with increasing temperature.} At low temperature there is a clear quasiparticle peak in the density of states. The quasiparticle peak gradually disappears in the bad metal regime which is centered around the QWL. The orange line is the density of states at the QWL.
The data are shown for the fixed chemical potential $\mu=3.5$ and $U=4$, which corresponds to roughly 15 \% doping.}
\label{fig_dos_evolution}
\end{figure}

We find that much better results are obtained by performing MEM on the spectral function
\begin{equation}
A(\varepsilon,i\omega_n)=\frac{1}{i\omega_n+\mu-\varepsilon-\Sigma(i\omega_n)}.
\end{equation}
The self-energy is then easily extracted from the real-axis result
\begin{equation}
\Sigma(\varepsilon; \omega)=\omega+\mu-\varepsilon-A^{-1}(\varepsilon, \omega).
\end{equation}
This procedure should in principle yield the same self-energy for any value of $\varepsilon$, but in practice this is not found to be the case. However, a good estimate of $\Sigma(\omega)$ is obtained by averaging the results of each continuation, i.e. 
\begin{equation}
\Sigma(\omega)=\frac{1}{N}\sum_{i=1}^{N}\Sigma(\varepsilon_i; \omega).
\end{equation}
Similarly, one could first calculate the Green's function
\begin{equation}
G(\omega)=\int d\varepsilon \rho_0(\varepsilon) A(\varepsilon,\omega)
\end{equation}
and then get the self-energy from the DMFT self-consistency. In practice, we have used 40 values of $\varepsilon$, equally spaced within the energy range of the non-interacting band, and found that the systematic and numerical error of MEM gets canceled by the averaging. We have found that in this approach, physically meaningful solutions are always obtained, results are more consistent and have less numerical noise, but at the expense of performing a much larger number of analytical continuations. Where available, we cross-checked our results with the findings in Ref.~\onlinecite{Deng2012SN} where the analytical continuation is performed via Pade approximant on the high-precision CTQMC data, and found very good agreement.

Given the self-energy on the real axis $\Sigma(\omega)$, the optical conductivity of the system can be calculated using the Kubo formula \cite{Georges1996SN}
\begin{equation} \label{realaxisres}
\sigma(\omega) \! = \! \sigma_o \! \int \! \! \int \!  d\varepsilon d\nu \Phi(\varepsilon) A(\varepsilon,\nu) A(\varepsilon,\nu+\omega) \frac{f(\nu)-f(\nu+\omega)}{\omega}, 
\end{equation}
where $ A(\varepsilon,\nu)=-(1/\pi)\mbox{Im}(\varepsilon+\mu-\nu-\Sigma(\varepsilon))^{-1}$, $f$ denotes the Fermi function, $\Phi(\varepsilon) = \Phi(0) [1-(\varepsilon/D)^2]^{3/2}$, and $\sigma_o=2\pi e^2/\hbar$. We present the resistivity results in the units of $\rho_{_{\mathrm{MIR}}}=\hbar D/e^2\Phi(0)$, consistently with Ref.~\onlinecite{Deng2012SN}.
We have calculated the resistivity $\rho_{_{\mathrm {DC}}}=\sigma^{-1}(\omega \rightarrow 0)$ in the whole $(\mu,T)$ plane. In Supplementary Figure \ref{fig_lines_and_doping} we show the lines of constant resistivity and constant doping in the $(\mu,T)$ plane. An example of the evolution of the density of states with temperature is shown in Supplementary Figure \ref{fig_dos_evolution}.

\section{V. Charge compressibility}

The QWL is defined in purely thermodynamic terms, from the free energy functional, and as such can be examined for any model. In fact, it does not even require the presence of a first order transition line with finite $T_c$. It is therefore important to explore physical properties other than the resistivity along and near the QWL. It is striking that the charge compressibility is nearly constant along the QWL, and has intermediate value between the one in almost incompressible Mott insulator and Fermi liquid, see Fig.~6.

\vspace*{.5cm}

\begin{figure}[h!]
\includegraphics  [width=3.2in]{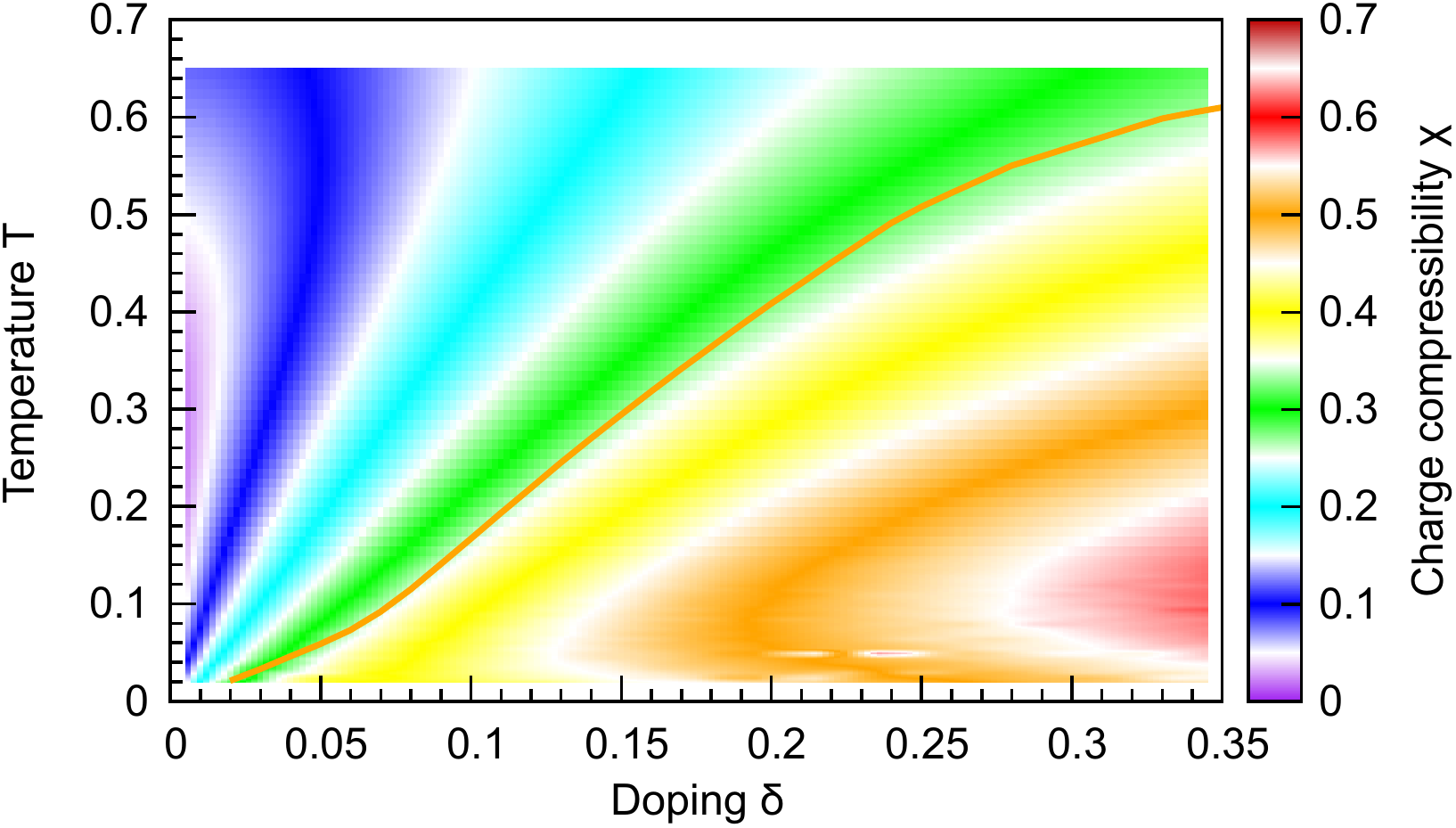}
\caption{\textbf {Color plot of the charge compressibility has the "fan-like" form, as generally expected for quantum criticality.} The compressibility is approximately constant along the QWL.}
\label{fig_comressibility}
\end{figure}

\section{VI. Boundaries of the QC scaling region}

\begin{figure*}
\includegraphics  [width=6.4in]{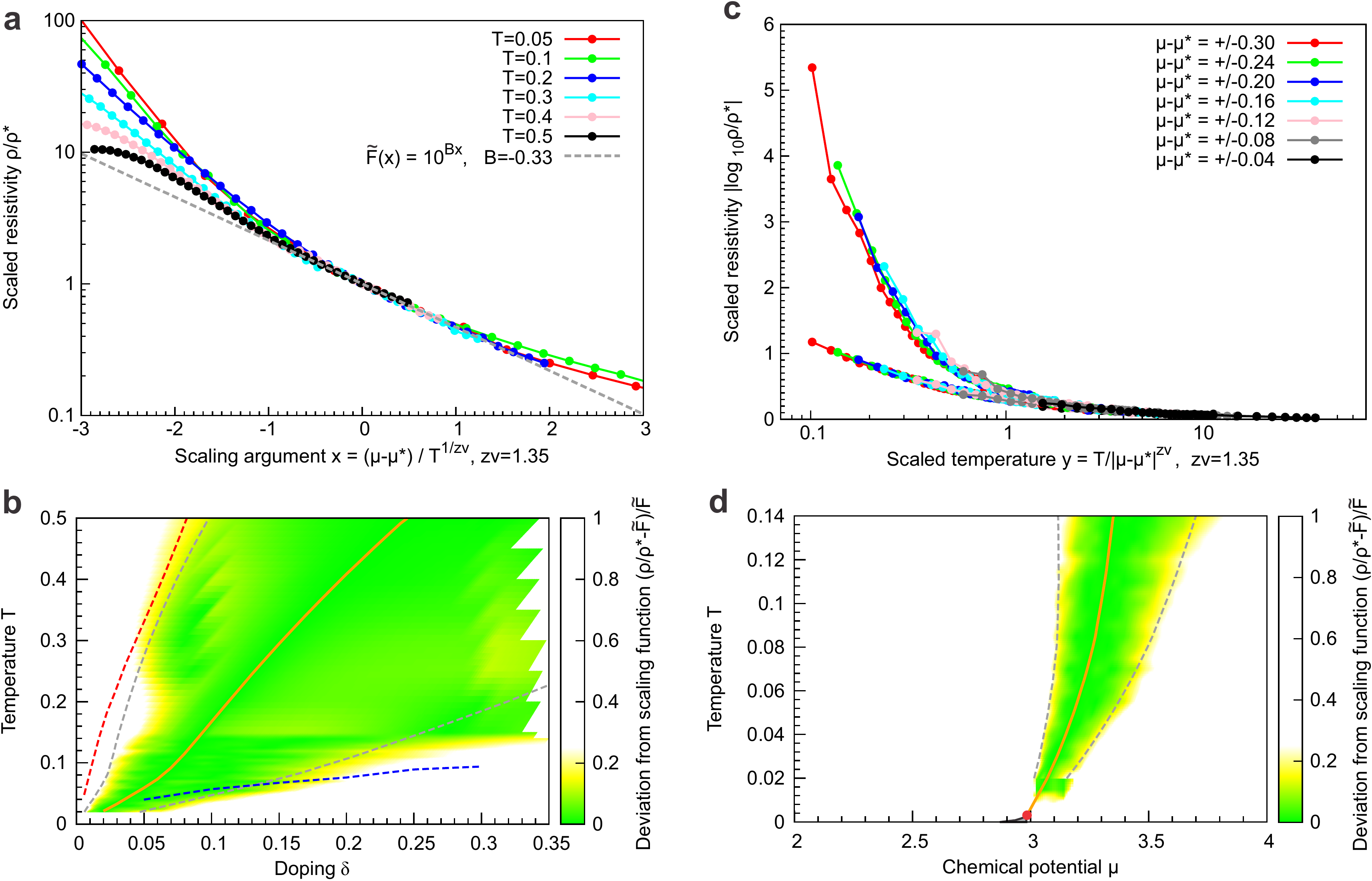}
\caption{\textbf{The extent of the scaling region.} (a) The DMFT data are plotted as a function of the scaling argument $x$ to obtain a fit for the scaling function. The range of x where DMFT data points fall on a single, well defined curve can be used as an estimate of the scaling region. (b) Between $x=-1.0$ and $x=1.5$ (gray dashed lines), the relative error of the scaling formula is below 10\%. The boundaries of the scaling region coincide with the $\mu=3.0$ line (dashed red) and the knee-like feature in resistivity $\rho(T)|_\delta$ which marks the boundary of the linear resistivity bad metal region (blue line). (c) The mirror symmetry is found where the two branches of $|\log F(y)|$ coincide. (d) The scaling region in the $\mu-T$ plane; the scaling is valid for $T \gtrsim 4T_c$.}
\label{scaling_boundaries}
\end{figure*} 

Around the quantum Widom line the resistivity is well approximated by a function of only one argument, such that
\begin{equation}\label{mu_scaling1}
\rho(\mu,T)= \rho^*(T) F\left(\frac{T}{d \mu ^{z\nu}}\right),
\end{equation}
where $\rho^*(T)=\rho(\mu^*(T),T)$ is the resistivity along the QWL. This behavior, typical for the quantum criticality, is shown in Fig.~2 in the Main Text, where the resistivity curves are collapsed on the metallic and insulating branch. The explicit form of the scaling function can be obtained from an equivalent scaling equation
\begin{equation}\label{mu_scaling2}
\rho(\mu,T)= \rho^*(T) \tilde{F}\left(\frac{d \mu}{T^{1/z\nu}}\right).
\end{equation} 
with the advantage of $\tilde{F}(x)$ being a smooth analytical function in $x$. Then,
the scaling function $\tilde{F}(x)$ can be obtained by plotting the DMFT resistivity data versus the argument $x=\frac{d \mu}{T^{1/z\nu}}$ and performing a numerical fit. This is shown in Supplementary Figure \ref{scaling_boundaries}a. $\tilde{F}(x)$ is approximately linear on the logarithmic scale which implies that $\tilde{F}(x)\approx10^{Bx}$, where $B\approx -0.33$. This analytical form is consistent with the mirror symmetry of the scaling formula near the QWL, $\tilde{F}(x)=1/\tilde{F}(-x)$. We can see that the scaling region goes beyond the mirror symmetry of the scaled resistivity curves, especially on the metallic side of the QC region. 

The scaling region can be estimated from the color plot of the relative error $r=|\rho_{DMFT}-10^{Bx}|/\rho_{DMFT}$, which is shown in Supplementary Figure \ref{scaling_boundaries}b. The boundaries of the QC scaling region defined by $r<10\%$ are shown with gray dashed lines and correspond to the values $x_{\mathrm{min}}=-1.0$ and $x_{\mathrm{max}}=1.5$. Note that they coincide with the $\mu=3.0$ line (red dashed; it corresponds to chemical potential being at the lower edge of the upper Hubbard band), and the knee-like feature in $\rho(T)|_\delta$ curves (blue dashed; it corresponds to the boundary of the linear resistivity bad metal region). It is obvious from this plot that the QC scaling region completely matches the region of typically bad metallic temperature dependence of the resistivity.

The boundaries of the QC scaling region can alternatively be estimated simply by looking at Fig.~\ref{scaling_boundaries}a and observing the maximum and minimum values of $x$ for which the DMFT results fall on a single well defined curve. This yields $x_{\mathrm{min}}=-1.0$ and $x_{\mathrm{max}}=1.5$. These lines are also shown in Fig.~\ref{scaling_boundaries}b (gray dashed) and are in good agreement with the independent estimate based on relative error $r$. 

Finally, the region of mirror-symmetry can be estimated by plotting the DMFT resistivity data $|\log\frac{\rho}{\rho^*}|$ as a function of $y=T/d \mu^{z\nu}$ (shown in Fig.~\ref{scaling_boundaries}c) and observing the lowest $y$ at which the two branches of data are found to coincide. This analysis yields $y_{\mathrm{min}}=1=|x_{\mathrm{min/max}}|^{-1}$, in agreement with other approaches.

\vspace*{12cm}

\pagebreak

\end{document}